Population pressure and global markets drive a decade of forest cover change in Africa's Albertine Rift

S.J. Ryan[a,b,c,d*], M. Palace[e,f], J. Hartter[g], J.E. Diem[h], C.A. Chapman[i], and J. Southworth[a]

[a]Department of Geography, University of Florida, Gainesville, FL 32601 USA *sjryan@ufl.edu

352-294-5955

[b]Emerging Pathogens Institute, University of Florida, Gainesville, FL 32601 USA

[c]Center for Global Health and Translational Science, Department of Microbiology and Immunology, Upstate Medical University, Syracuse, NY 13210 USA

[d]School of Life Sciences, University of KwaZulu-Natal, Durban, South Africa

[e]Institute for the Study of Earth, Oceans, and Space, University of New Hampshire, Durham, NH 03824-3525 USA

[f]Department of Earth Sciences, University of New Hampshire, Durham, NH 03824-3525 USA

[g]Environmental Studies Program, University of Colorado, Boulder, CO 80309 USA

[h]Department of Geosciences, Georgia State University, Atlanta, GA 30303 USA

[i]Department of Anthropology and School of Environment, McGill University, Montreal, Quebec, H3A 2T7 Canada

**Abstract**

Africa's Albertine Rift region faces a juxtaposition of rapid human population growth, while being a biodiversity hotspot. Using satellite-derived continuous forest cover change data, we examined national socioeconomic, demographic, and agricultural production data, and local demographic and geographic variables to assess multilevel forces driving significant local forest cover loss and gain outside protected areas during the first decade of this century. Because the processes that drive forest cover loss and gain are expected to be different, and both are of interest, we constructed models of change in each direction. Although forest cover change varied by country, national level population change was the strongest driver of forest loss rate for all countries – with a population doubling predicted to cause 2.06% annual cover loss, while doubling tea production was predicted to cause 1.90%. The rate of forest cover gain was associated positively with increased production of the local staple crop cassava, but negatively with local population density and meat production, suggesting production drivers at multiple levels mitigate reforestation. We found a small, but significant, decrease of forest cover loss rate with increasing distance from protected areas, supporting studies suggesting higher rates of landscape change near protected areas. While local population density mitigated the rate of forest cover gain, cover loss also correlated to lower local population density, an apparent paradox, but consistent with findings that larger scale forces outweigh local drivers of deforestation. This implicates demographic and market forces at national and international scales as critical drivers of change, calling into question the necessary scale of forest protection policy in this biodiversity hotspot. Use of MODIS continuous forest cover products, and the study of both forest cover gain and forest cover loss, add a dynamic component to more traditional static and uni-directional studies, significantly improving our understanding of the landscape processes and drivers at work.

1. **Introduction**

The Albertine Rift region of Africa, comprising parts of Uganda, Tanzania, Burundi, Rwanda, the Democratic Republic of Congo (DRC), and Zambia, is a biodiversity hotspot (Plumptre et al. 2003, 2007). The juxtaposition of some of the highest human population growth rates and densities in the world, and richest conservation areas, also make it one of the world's most vulnerable conservation-poverty hotspots (Fisher and Christopher 2007). One of the biggest concerns about the increasingly isolated protected areas in the region is the pressure exerted by deforestation and demand for arable land by a rapidly increasing and dense rural human population outside of protected areas (Hartter and Southworth 2009, Hartter et al. 2011). While rural poverty and local population pressure are often cited as reasons for deforestation in developing countries (Sassen et al. 2013), reviews suggest that the evidence does not support it (Rudel 1996), and instead, multiple levels of socioeconomic and political forces at local, national and global levels operate to determine deforestation (Lambin et al. 2001, Geist and Lambin 2002, Lambin and Meyfroidt 2010, 2011, Meyfroidt et al. 2010).

The forest transition theory (FT), predicts that as GDP increases, corresponding to development, deforestation due to agricultural conversion accelerates, then plateaus, and finally starts to decrease (Walker 1987, Mather 1992, Rudel et al. 2002, Lambin and Meyfroidt 2010, 2011, Meyfroidt et al. 2010). Rather than increasing poor rural populations driving deforestation, large scale human migration patterns associated with urbanization may deplete the rural population, leaving marginal lands to recover, leading to areas of reforestation (Aide and Grau 2004, Grau and Aide 2008, Aide et al. 2013). However, an increasingly urban population may also place higher demand on rural fuelwood economies, because cities require charcoal, which is brought in to markets (Mackenzie and Hartter 2013), creating a counterintuitive signal of population density and deforestation. Importantly, GDP itself does not capture the distribution of wealth, capital, or land tenure in developing countries. An extension of the FT theory is the rise of urban consumers, driving a demand for more expensive, higher protein diets -the nutrition transition (Popkin 2001); and the demand for meat production also increases pressure on the land. These FT theories and extensions have not been tested in Africa to our knowledge, and instead derive from studies in other tropical forested regions, such as the Amazon, where population density is an order of magnitude lower.

In a global analysis of drivers of forest loss between 2000 and 2005, DeFries et al. (2010), argued that urban population growth and agricultural trade were the biggest drivers of tropical forest loss. However, this was rebutted by Fisher (2010), who provided evidence that, for the 12 African countries in the analysis, deforestation was instead correlated to staple crop production, suggesting a rural domestic crop production link. Our study comprises a more recent set of data (2000-2010), and tests both these sets of drivers (domestic staple and export production). The two previous studies unfortunately only overlap with our countries in one case (DRC), precluding direct comparison, but we can draw parallels in interpretation. In the Albertine Rift, both national and sub-national population trends over the first decade of this century may also have been influenced by armed conflict, rural to urban migrations, climate impacts, and global market forces influencing national commodity and monetary flows (DeFries and Rosenzweig 2010).

The Albertine Rift has experienced considerable conversion of land to agriculture, from extremely small-scale multi-cropping subsistence agriculture, fragmenting and subdividing rural landscapes around protected areas, to cooperative farming for tea (Hartter and Southworth 2009). While tea cooperatives are simply smaller in size, and tend not to require the massive fertilizer inputs and tilling that, for example, U.S. industrial corn production uses, their establishment nonetheless results in substantial and lasting conversion of the landscape (Hartter and Southworth 2009, Hartter et al. 2011, Ryan and Hartter 2012, Ryan et al. 2015). Although tropical forest conversion for meat production is well known as an Amazonian frontier problem (Alves et al. 2009), in Africa, an increasingly affluent and urban population, driving higher meat consumption, likely impacts land cover and land use changes in the Albertine Rift.

We used a spatially random sample of 100,000 points across the Albertine Rift to analyze forest cover change (percent canopy cover), derived from moderate resolution satellite imagery (MODIS) from 2000-2010 at a 250m resolution. The use of this MODIS dataset allows for the development of a continuous forest cover change analysis, because it is not simply a binary description of forest presence/absence, and allows us to develop models of the rate of both forest cover loss (deforestation) and forest cover gain (reforestation). This is a significant improvement over unidirectional approaches, and allows us to review both deforestation and reforestation drivers and processes across this key landscape. Ten annual estimates of percent canopy cover were used to develop linear regressions for each point. We examined locations with significant rates of forest cover change (loss and gain) over the decade, and analyzed these in a multi-level model to simultaneously assess the impacts of local and national level drivers on local forest change. We examined the relationship between small-scale, local forest cover change and potential demographic, economic, and agricultural production drivers of deforestation at the local and national level for the six countries that constitute the Albertine Rift (Box 1).

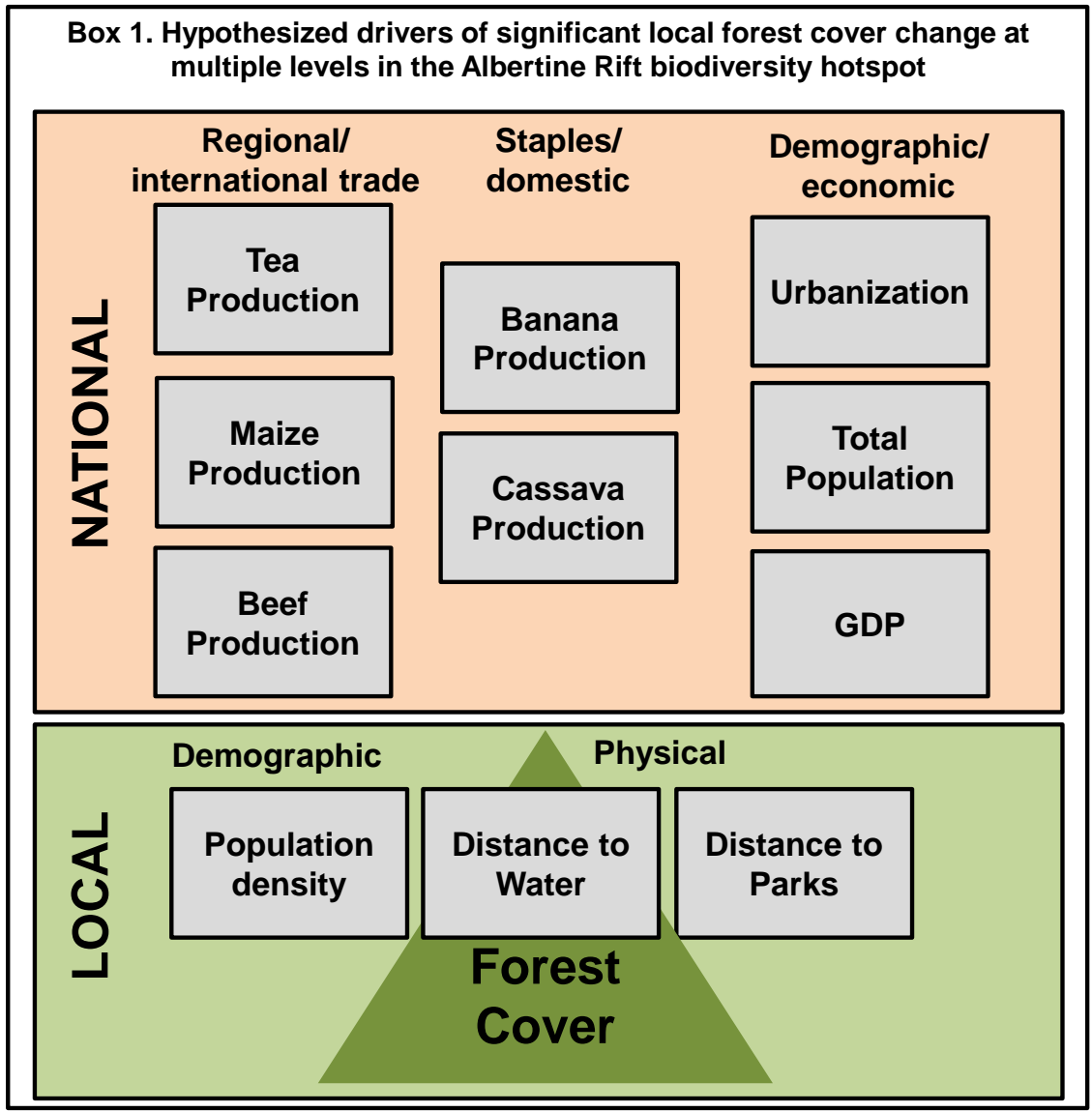

Box 1. Hypothesized drivers of significant local forest cover change at multiple levels in the Albertine Rift biodiversity hotspot

## 2. Materials and Methods

### 2.1 Data acquisition and processing

We sampled the MODIS data derived product, VCF (Vegetation Continuous Fields - MOD44B, 250 m resolution) using 100,000 spatially random points across the Albertine Rift. Regressions for each point were used to derive a rate of annual forest cover change between 2000 and 2010 using ten annual estimates of percent forest cover (Hansen et al. 2005). We only retained significant regressions (at $p \leq 0.05$, n = 8,175), and removed all points that fell inside protected areas. We divided the remaining points into two datasets: forest cover gain (positive slopes, n = 3,835), and forest cover loss (negative slopes n = 2,685). We note that using MODIS-derived forest cover change may obscure very small-scale change (less than 250 m) and that VCF often has annual fluctuations and errors. The advantage of this product over more simple binary forest/non-forest cover products (e.g. (Hansen et al. 2013)), is that this MODIS product allows for sub-pixel proportional estimates of tree crown cover and provides an estimate preserving a continuous variable of landscape heterogeneity. It is also important that we model both directions of forest cover change – so we can see direction of change and intensity of change and link these landscape processes to the drivers across multiple spatial scales.

The Albertine Rift contains more than 100 protected areas, and also many important lakes (Figure 1), which likely influence patterns of human settlement in the region. We controlled for the potential impact of large water bodies on human settlement by including distance to water as a random intercept in the multilevel model. Our 'local' per pixel population density change estimates were derived from the 100m resolution WorldPop (of which Afripop is a subset) dataset (Linard et al. 2012), aggregated to 250m resolution for 2010. We chose not to use a change estimate since the data underlying this product for at least one of the countries (Uganda, see Linard et al. (2012), supplementary information) do not permit this. Instead, we preferred to use a single snapshot of estimated local population density. Agricultural and market production measures (tea, maize, meat, bananas, cassava, in tonnes), for countries from 2001-2010 were acquired from the Food and Agricultural Organization (FAO) (faostat.org, *accessed Sept 10$^{th}$, 2015*), and demographic and economic variables for 2001 and 2010 (population, rural proportion, GDP) were acquired from the World Bank (data.worldbank.org, *accessed Sept 10$^{th}$, 2015*). These were included in models as proportional changes.

## 2.2 Analysis

We used multilevel models in R (version 2.13.2; packages 'lme4', 'arm', 'glmulti') to explore the roles of demographic and environmental change at the local, pixel level, and the additional roles of demographic and market forces at the national level. There are many languages used to describe multilevel, hierarchical, mixed effects, or nested models, which vary between and even within disciplines. For ease of explanation, we refer to this regression as a multilevel model, and describe the variance components as they appear in our specific case.

We used a basic two-level model to describe the significant slope of forest cover change, $y_{ij}$, at the pixel level ($i$), within country ($j$):

$$\text{Level 1: } y_{ij} = \beta_{0j} + \beta_{1j}(X_{1ij}) + \beta_{2j}(X_{2ij}) + ... + \beta_{nj}(X_{nij}) + e_{ij} \quad \text{eqn.1}$$

Where $X_{1...n}$ are predictor variables at the pixel level; $e_{ij}$ is the error term subsuming the independent error term for the intercept $\beta_0$ and the independent error of the regression coefficients $\beta_1$ to $\beta_n$, and the predictors $X_1$ to $X_n$. $\beta_0$ to $\beta_n$ are the regression coefficients, whose variation depends on explanatory variables at the country level, for example:

$$\text{Level 2: } \beta_{0j} = \gamma_{00} + \gamma_{01}Z_j + \mu_{0j} \quad \text{eqn. 2}$$

In which $\gamma_{00}$ is the intercept for the overall model of $\beta_0$, and $Z_j$ is the country-level predictor, with the residual error $\mu_j$ at the country level.

As we had multiple local and country level predictors, this can be summarized with $X$ taking subscript $p$ (1...P), and $Z$ taking $q$ (1...Q), as:

$$Y_{ij} = \gamma_{00} + \sum_p \gamma_{po} X_{pij} + \sum_q \gamma_{0q} Z_{qj} + \sum_p \sum_q \gamma_{pq} X_{pij} Z_{qj} + \sum_p \mu_{pj} X_{pij} + \mu_{0j} + e_{ij} \quad \text{eqn. 3}$$

As our hypotheses included predictors that are likely to be correlated, we mean-centered all variables (described below), examined variance inflation factors (VIFs) of the parameters, and kappa statistics for collinearity effects on the overall model, using the package 'mer-utls.R' ('mer-utls.R', https://github.com/aufrank/R-hacks.git, *accessed 09 September, 2015*).

For predictor variables available in 2000 and 2010, we used proportional changes in the indicators, and centered the variable. Centering predictors is useful for interpretation in multi-level models by allowing examination of relative change on the mean (average) property of a level at the higher level. Additionally, it tends to improve model convergence. Distance to parks and distance to water were continuous variables (in meters) that did not change over decade, so were divided by 10,000 (to create a 100km variable for model fitting), and centered.

We conducted predictor selection in a multi-model comparison using Akaike's Information Criterion (AIC) to select the best candidate set of variables (Burnham and Anderson 2002), using the R package 'glmulti'(Calcagno and de Mazancourt 2010). The base (or intercept only) model $y_{ij} = \beta_{0j} + e_{ij}$, was used to establish the structure in the data at the country level, to compare the impact on model fit of adding predictors at the two levels (pixel (*i*) and country (*j*)). We created baseline models for negative and positive slopes (cover loss and gain, respectively), and derived AIC values. We then stepped through two stages of predictor and factor addition: adding a random component to the model, country, then the distance to water, and then using model selection for the fixed component on the remaining predictors constraining the number of variables to 4 to avoid overfitting, at the country and pixel level.

We included national production estimates of meat, maize, banana, cassava, and tea. We also included GDP, national total population, the proportion of rural population derived from FAO and World Bank data, local population density, and the distance to nearest water body and protected area as independent variables. In each stage, model improvement over the previous was assessed, with the criteria of 'improvement' at ΔAIC ≥ 2 (Anderson et al. 2001, Burnham and Anderson 2002).

For the best model fits, we report the significance of parameters using t-tests, understanding that our large sample size and relatively few estimated parameters obscured the uncertainty about estimating degrees of freedom (DF), which would exceed 500, often the point of reported convergence of the critical value at 1.96 for $\alpha = 0.05$.

## 3. Results

There was considerable variation across the Albertine Rift countries in the rate of significant local forest cover change in our random sample from 2000-2010 [ANOVA: cover loss: F=108.48, df=5, p<0.0001; cover gain: F=11.44, df=5, p<0.0001]. The mean and range of rates of gain and loss were 1.28% [0.13-6.97] and -1.98% [-0.19- -7.20] respectively, suggesting higher rates of loss than gain from 2000-2010 (Figure 2). Within our random sample, there were more pixels with significant forest cover gain (positive slopes, n=3,835), than forest cover loss (n=2,685), indicating that both processes occurred during the decade, and that while the rate of significant loss was greater than the rate of gain, more pixels were significantly gaining cover.

**3.1 Model Fits**

The simple model of forest cover change by country was our base model (Gain: AIC = 8,745; Loss: AIC = 7, 147). Adding the distance-to-water component to the random effect improved model fit considerably (Gain: AIC = 8,433, ΔAIC = 309; Loss: AIC =7, 065, ΔAIC = 81).

We then performed model selection on our fixed component, constraining the number of variables to 4 to avoid overfitting, and arriving at the top models, with Gain: AIC = 8,418 (ΔAIC = 15); and Loss: AIC = 7,044 (ΔAIC = 21), demonstrating considerable improvement in support. The best-fit model for forest cover loss included national tea production, national population change, local population density, and distance to protected areas, as drivers of forest cover loss (Table 1). The best-fit model for forest cover gain included national banana and cassava production, meat production, and local population density (Table 1). We show the trajectories of these drivers across prediction space in Figure 3.

**3.2 Model stability**

When we examined the final models for collinearity, using the Kappa statistic ('mer-utls.R', https://github.com/aufrank/R-hacks.git), we obtained values of $\kappa = 1.72$ for the loss model, and 1.88 for the gain model, suggesting low collinearity. We calculated variance inflation factors (VIFs) to assess all predictors, and found that our predictors all fell below 1.5, indicating little inflation and thus we could be confident in our estimates of model fit (Gelman and Hill, 2006).

4. **Discussion**

An analysis of significant forest cover change outside of protected areas in the first decade of the 21$^{st}$ century across the Albertine Rift suggests that a dynamic process of forest cover gain *and* forest cover loss is occurring on the landscape, with different drivers emerging at local and national scales, and signals of higher *rates* of forest cover loss, but greater *areas* of forest cover gain. This again serves to highlight the advantages of using both the continuous MODIS derived forest cover product and modelling both cover loss and forest cover gain, to more fully understand our landscape and its patterns and drivers of change.

National level population change was the biggest driver of forest loss rate; a doubling in population is predicted to cause 2.06% forest loss per year at the local level. In Uganda, which has a population growth rate of 3.24% (Central Intelligence Agency, 2015), population doubling would occur in just over 20 years. The increase in population across the Rift will put ever more pressure on remaining forests for fuelwood, timber for housing, and other resources, in addition to continued land clearing for agriculture.

At the local level, an increase in population density was associated with lower rates of cover loss, but was found to mitigate for forest cover gain. While this seems contradictory, it supports recent work suggesting that local population pressure is not a main driver of deforestation (DeFries et al. 2010), yet reversion to forest is likely slowed by the need for arable land. In sub-Saharan Africa, rural livelihoods depend on land and natural resources (Abulu and Hassan 1998); increasing population growth and a trend towards middle-income economies is placing ever-increasing pressures on land use and natural resources (Hartter and Ryan 2010).

The increase in production of both tea and meat at the national level were found to be drivers of cover change, tea was responsible for 1.90% annual forest loss, while meat corresponded to -0.71% annual forest gain, with a doubling in production. These two markets may operate in different ways to exert this change. Tea is largely produced for global export markets (Chapagain and Hoekstra 2007), exported to, and departing from Kenya (FAOSTAT.org, *accessed Sept 10$^{th}$, 2015*), while meat produced in the countries of the Rift is largely traded within sub-Saharan Africa (FAOSTAT.org, *accessed Sept 10$^{th}$, 2015*). While tea production was initially a holdover from colonial land ownership and markets oriented toward supporting Great Britain, it clearly continues to be a major market force in the region. Tea transforms the landscape by introducing a perennial crop, rather than seasonal subsistence level crops. Tea requires larger land parcels, added transportation infrastructure, and fuelwood for drying. It also transforms the scale of economy, as tea businesses may induce cooperatives, requiring a steady labor force. Potentially of significance for biodiversity maintenance, in a region where almost no chemical fertilizers are used, tea is heavily fertilized (Freeman and Omiti 2003), adding potential ecotoxicological impacts on landscape health. Meat production, on the other hand, appears to be a largely regionally traded commodity, and is a signal of the rising demand for meat in affluent, more urbanized systems – this coupled forest-nutrition transition.

In contrast to these trade commodities, we found that forest cover gain was influenced by staple crop production. Following Fisher (2010), in ascertaining the role of domestic agriculture in the form of staple crop production, we found there was a small but positive association between forest cover gain and cassava production, suggesting that perhaps this staple is not a primary driver displacing forest cover in the region despite its increasing production levels throughout the decade in all countries considered. In contrast, we found that banana production was negatively correlated with forest cover gain, indicating that perhaps rural lands used to produce bananas slow the process of reforestation. Unfortunately, we did not have the resolution of data to ascertain what type of banana was involved in this production measure – banana is a term used for both starchy cooking staple types of banana, and the sweet banana, which may be part of a regional or even international production market.

We found a small (0.4% per 100km per year), but significant, trend of decrease in forest cover loss rate as the distance from protected areas increased, across six countries. This finding aligns with previous work suggesting that human density is higher near protected areas (Wittemyer et al. 2008), but may be more indicative not of settlement patterns, rather highly dynamic land-use in the wake of park border establishment (Ryan and Hartter 2012, Ryan et al.

2015). The process of increasing isolation of remnant forest cover continues to challenge protected area management, indicating a need to understand and manage drivers of deforestation.

As protected area management and human livelihoods continue to exert impacts on the landscape of the Albertine Rift, it is important to identify the driving forces behind forest cover gain and loss, as this information is needed to guide conservation strategies. We saw that different drivers of cover change are acting on forest cover. We found that national production markets for both meat and tea are significant drivers of both direct loss, and a reduction in cover gain, respectively, but serve different scales of consumers. Tea is meeting a global demand, whereas meat production is meeting regional dietary shifts towards higher meat consumption in more affluent urban areas. We saw that staples, cassava and banana, which were the proxy for domestic production and consumption – rural small scale agriculture – yielded a mixed signal, but were important to the processes driving cover gain over the decade. Population growth and local population density were also very important drivers, and are more complicated to address as conservation policy than production markets. Targeting policy at the appropriate scale to protect forest cover, and identifying the scope of population pressure on remaining fuelwood supplies is fundamental to protecting this biodiversity hotspot.

This study indicates that the coupled forest transition and development spectrum are not restricted to Latin America and the Caribbean, and point to the need to understand how development, land tenure legislation, and population growth will proceed in conjunction with forest and timber trade policies at multinational scales. While African deforestation rates have been found to be slower than other tropical forested areas (DeFries et al. 2010, Fisher 2010), and most countries in this analysis are far lower on development trajectories than those of primary focus for FT theory (e.g. Brazil), our findings indicate emergence of similar market and consumption drivers, at multiple scales. Most importantly, our methods of using multi-level modeling in association with a continuous measure of forest cover change, allowing us to understand both forest cover loss and forest cover gain, as well as intensity versus area of these changes, is a significant step forward in understanding the drivers of change across such important, biodiversity hotspots.

**Acknowledgments**

This research was supported by a National Science Foundation Coupled Natural-Human Systems Exploratory grant (NSF-EX: 1114977). Nick Dowhaniuk provided graphic support and Christina Czarnecki provided GIS support.

**Author Contributions**

SJR conceived the framework, conducted statistical analysis; MP processed data and contributed to analysis; SJR, MP, JED, CAC, JH and JS wrote and edited the manuscript.

**Table and Figures**

Table 1. Parameter estimates and significance (*p*-value) for the best-fit model of significant deforestation across the Albertine Rift, Africa.

Figure 1. The Albertine Rift region of Africa. Note lakes and protected areas. This figure was produced using ArcGIS 10 (*ArcGIS 10* 1999).

Figure 2. Significant rates of annual forest cover change 2001-2010 (%/yr of (a) loss and (b) gain), in six countries, outside protected areas and water bodies in the Albertine Rift, from a geographically random sample of 100,000 points. Boxes encompass 25-75% quartiles of the mean.

Figure 3. The individual partial effects of the predictors in the top models (with 95% confidence interval bands in grey), for forest cover gain (a-d) and loss (e-h), as a function of a. cassava production change; b. banana production change; c. meat production change; d. local population density; e. distance from nearest park (in 100km units); f. tea production change; g. national population change; and h. local population density.

Table 1:

| Model | Variable | Estimate* | p-value |
|---|---|---|---|
| *Cover loss* | Total Population | -2.06 | <0.0001 |
| | Tea | -1.90 | <0.0001 |
| | Park Distance (100km) | 0.394 | <0.0001 |
| | Population Density | 0.026 | <0.0001 |
| *Cover gain* | Banana | -0.049 | <0.001 |
| | Cassava | 0.25 | <0.0001 |
| | Meat | -0.707 | <0.0001 |
| | Population Density | -0.009 | <0.05 |

*% annual rate of forest cover change

**Figure 1**

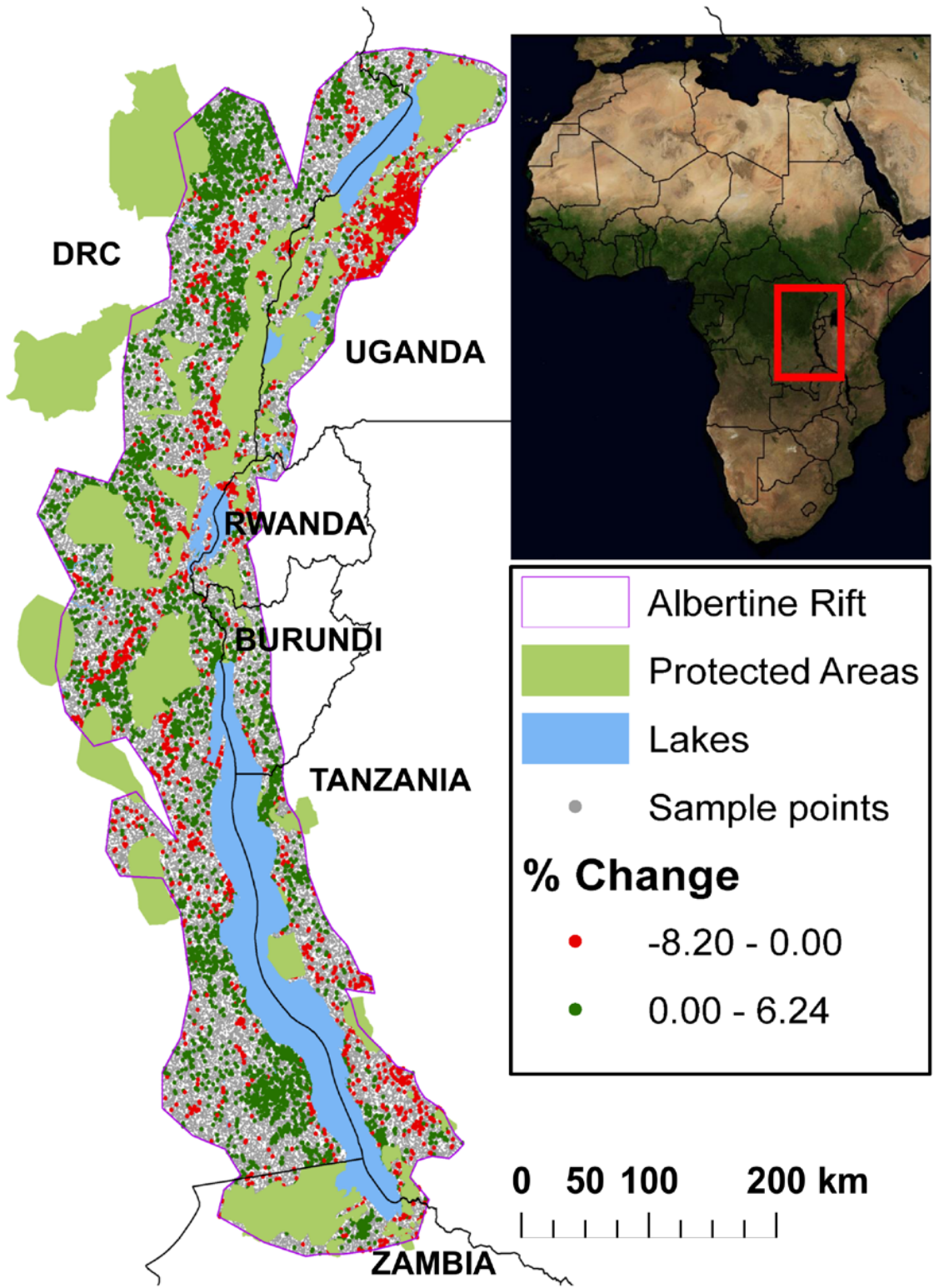

**Figure 2**

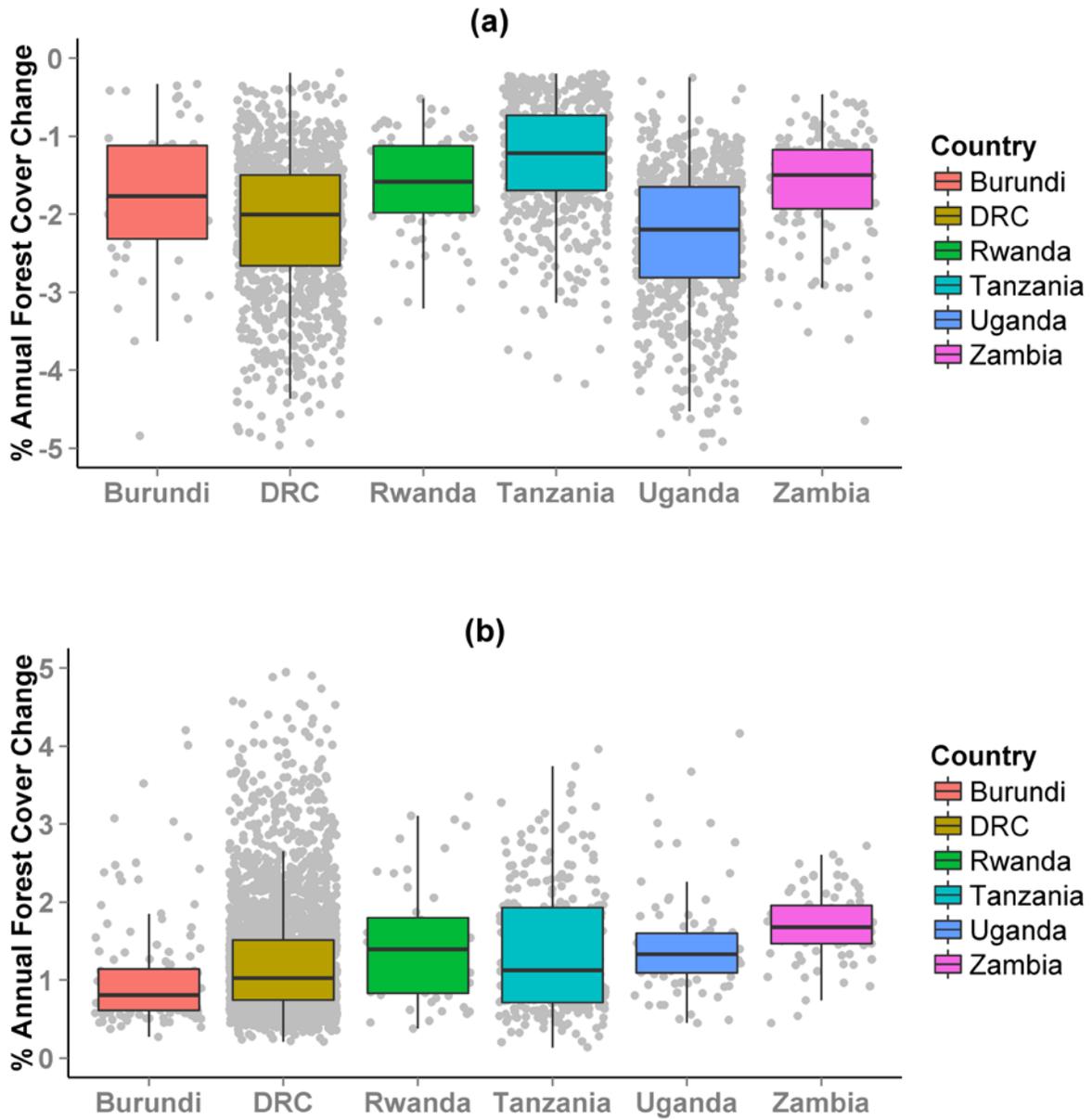

**Figure 3**

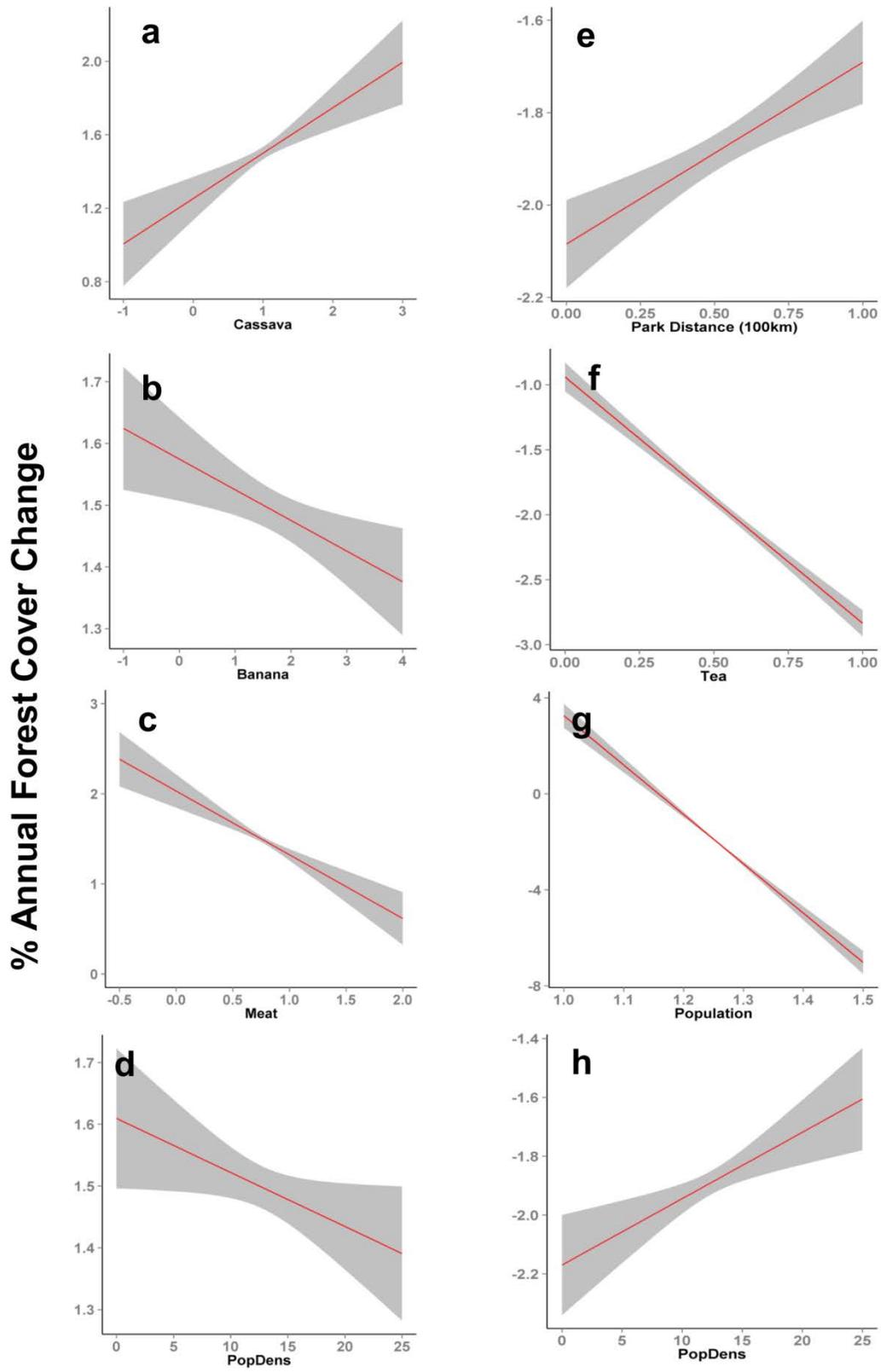